\documentclass[12pt]{article}   	
\usepackage{titlesec}
\usepackage[text={6.5in,8.5in}]{geometry}                		
\usepackage{graphicx}				
\usepackage{amssymb}
\usepackage{color, hyperref}
\usepackage{xcolor}
\usepackage{sectsty}
\usepackage{feynmp-auto}

\definecolor{grey}{rgb}{0.3,0.3,0.3}
\definecolor{maroon}{rgb}{0.6,0.1,0.2}

\chapterfont{\color{maroon}}  
\sectionfont{\color{maroon}}
\subsectionfont{\color{grey}}

\newcommand{\ttt}[1]{\texttt{#1}}

\newcommand{\maroon}[1]{\textcolor{maroon} {#1}}

\begin{document}
\begin{center}
\maroon{\LARGE \textbf{Collider signatures for dark matter and long-lived particles with} \textsc{Pythia 8}} 
\vskip 0.1in
Nishita Desai\footnote{nishita.desai@umontpellier.fr} \\
Laboratoire Univers et Particules de Montpellier, \\CNRS-Universit\'e de Montpellier, 34095 Montpellier, France 

\end{center}

\begin{abstract}
We describe the implementation of four models for production of dark matter or associated particles at the LHC based on the simplest extensions of the Standard Model. The first kind of models include dark matter production via s-channel mediators.  This includes production in association with a jet for a vector boson ($Z^\prime$) or scalar ($S$) mediator as well as mono-higgs production via associated $hZ^\prime$ production.  We implement the simplest t-channel mediator in the form of a scalar with leptonic quantum numbers and completely generic Yukawa couplings to the dark matter fermion and a right-handed SM lepton.  Finally, we implement a generalised model of mixed dark matter where the dark matter is a mixture of an SU(2) singlet and N-plet.  We find that the last two models are also ideally suited to study the production of a range of long-lived particle signatures.  We illustrate this by showing the complementarity of the limits from various long-lived searches for a simple benchmark case.
\end{abstract}

\vskip 0.1in
\hrule
\vskip 0.2in

\tableofcontents


\pagebreak

\section{Introduction}

With the higgs scalar discovered at the LHC looking increasingly Standard Model (SM)-like~\cite{Khachatryan:2016vau}, and a lack of signal in searches for well-motivated models like supersymmetry, the search strategy at the LHC has shifted to a more bottom-up viewpoint focussing on covering all possible signatures of new particles.  This has prompted increasing interest in ``model independent'' searches for dark matter and other exotic final states, including new long-lived particles and dedicated working groups have been working on compiling a comprehensive search strategy~\cite{Abdallah:2015ter, Abercrombie:2015wmb}.  In this note, we describe the Pythia\,8~\cite{Sjostrand:2014zea} implementation\footnote{Available in the next public release of Pythia\,8.} of simplified models that allow the simulation of a range of theoretically well-motivated dark matter and long-lived particle scenarios.

We implement two models with s-channel mediators to a fermionic dark matter particle.  The first of these is a vector boson mediator ($Z^\prime$ or dark photon) with vector or axial-vector couplings.  The axial-vector model is a particularly well-motivated extension of the Standard Model as it averts strong bounds from direct detection experiments and will indeed be first observable at the LHC~\cite{Lebedev:2014bba}.  We implement both resonant production and production associated with one jet to cover the primary signatures.  The second model is a scalar s-channel mediator (also with possible pseudo-scalar couplings), as is commonly required by higgs-portal models~\cite{Patt:2006fw}.  Such scalars would also arise in the presence of extra gauge sectors which are spontaneously broken and therefore predict extra higgses.  Again, we implement resonant production as well as production in association with one jet.  Finally, we include production of a single higgs (i.e. a ``mono-higgs"\cite{Carpenter:2013xra}) via associated production of the higgs with the dark photon.

In the second part, we implement two models for producing charged partners of dark matter via Drell-Yan production.  Such partners are predicted by dark matter models where the correct relic density is achieved by co-annihilation with a charged partner~\cite{Ellis:1999mm, Mizuta:1992qp}  or by non-thermal production via freeze-in~\cite{Hall:2009bx}.  The simplest extension is by adding a scalar that has Yukawa couplings between SM and DM, and by choosing the scalar to have only leptonic charges, we can evade current direct detection bounds.  

The last model we implement is a generalisation of the supersymmetric dark matter sector by postulating that the dark matter is a mixture of a singlet and SU(2) N-plet fields (or ``next to minimal dark matter'' ~\cite{Bharucha:2018pfu, Bharucha:2017ltz}), but without imposing supersymmetry which requires the coupling of the dark sector gauginos to be the same as gauge couplings.  Choosing $N=2$ is similar to bino-higgsino dark matter~\cite{Gogoladze:2010ch} of supersymmetry or singlet-doublet model~\cite{Cohen:2011ec} whereas $N=3$ is similar to bino-wino dark matter~\cite{Baer:2005jq}.  We also include the option to choose $N=5$, motivated by the minimal dark matter~\cite{Cirelli:2005uq} idea which says the neutral component of a 5-plet is naturally stable because it is not possible to write a gauge invariant, renormalisable operator that can determine it's decay.  By choosing masses or the mixing parameter, it is also possible to simulate production of pure doublet, triplet or quintuplet states.

\section{Models and implementation}

The four models implemented here can be used to study a range of prompt and long-lived signatures at the LHC, listed in table~\ref{tab:sig} below.  The prompt signatures include mono-jet~\cite{Aaboud:2017phn}, dijet resonance~\cite{Aaboud:2017yvp, Sirunyan:2018xlo}, mono-higgs~\cite{Aaboud:2016obm, Sirunyan:2017bsh} and dileptons + missing energy (MET)~\cite{Aaboud:2018jiw, Sirunyan:2018nwe}.  The long-lived signatures possible include charged tracks~\cite{Chatrchyan:2013oca}, disappearing tracks~\cite{Aaboud:2017mpt, Sirunyan:2018ldc}, and displaced leptons~\cite{Khachatryan:2016unx} or vertices~\cite{Aaboud:2017iio}. The full list of processes and parameters can be found in the accompanying appendices.

\begin{table}[h!]
\begin{center}
\begin{tabular}{|c|cccc|}
\hline
{\bf Signature} & {\bf Dark photon} & {\bf Scalar} & {\bf Scalar} & {\bf n-plet singlet}  \\
& or $\mathbf{Z}^\prime$ & {\bf (neutral)} & {\bf (charged)} & {\bf mixed} \\
\hline
& monojet & monojet & di-leptons + MET & di-leptons + MET \\
Prompt & dijet resonance & dijet resonance & & \\
& mono-Higgs & & & \\
\hline
& & &  charged tracks &  charged tracks \\
Long- & &  & disappearing track & disappearing track \\
lived & & & displaced leptons & displaced leptons \\
& & & & displaced vertex \\
\hline
\end{tabular}
\end{center}
\caption{\label{tab:sig} Prompt (top) and long-lived (bottom) signatures possible from the implemented models. }
\end{table}

\subsection{Vector and scalar s-channel mediators}
\begin{figure}[t]
\begin{fmffile}{monoH}
\vskip 0.2in
\begin{center}
\begin{fmfgraph*}(80,60) 
\fmfset{arrow_len}{2mm}
\fmfleft{i1,i2}
\fmfright{o1,o2,o3}
\fmflabel{$\chi$}{o3}
\fmflabel{$\chi$}{o2}
\fmflabel{$q$}{i1}
\fmflabel{$q^\prime$}{i2}
\fmflabel{$g$}{o1}
\fmf{fermion}{i1,v1,v2,i2}
\fmf{curly}{v1,o1}
\fmf{photon, label=$Z^\prime$}{v2,v3}
\fmf{fermion}{o2,v3,o3}
\end{fmfgraph*}
\hspace{1.5cm}
\begin{fmfgraph*}(80,60) 
\fmfset{arrow_len}{2mm}
\fmfleft{i1,i2}
\fmfright{o1,o2,o3}
\fmflabel{$\chi$}{o1}
\fmflabel{$\chi$}{o2}
\fmflabel{$q$}{i1}
\fmflabel{$q^\prime$}{i2}
\fmflabel{$h$}{o3}
\fmf{fermion}{i1,v1}
\fmf{fermion}{i2,v1}
\fmf{photon, label=$Z^\prime$}{v1,v2}
\fmf{scalar}{v2,o3}
\fmf{photon, label=$Z^\prime$}{v2,v4}
\fmf{fermion}{o1,v4}
\fmf{fermion}{v4,o2}
\end{fmfgraph*}
\hspace{1.5cm}
\begin{fmfgraph*}(80,60) 
\fmfset{arrow_len}{2mm}
\fmfleft{i1,i2}
\fmfright{o1,o2}
\fmflabel{$\chi$}{o1}
\fmflabel{$\chi$}{o2}
\fmf{curly,tension=1.0}{i1,v1}
\fmf{curly,tension=1.0}{i2,v2}
\fmf{fermion,tension=1.0}{v1,v2,v3,v1}
\fmf{scalar,label=$S$,l.side=left}{v3,v4}
\fmf{fermion,tension=1.0}{o1,v4,o2} 
\end{fmfgraph*}
\end{center}
\end{fmffile}
\caption{From left to right ({\it a}\/) Dark photon + mono-jet, ({\it b}\/) Mono-higgs and ({\it c}\/) Higgs-portal production of DM. \label{fig:s-chan}}
\end{figure}
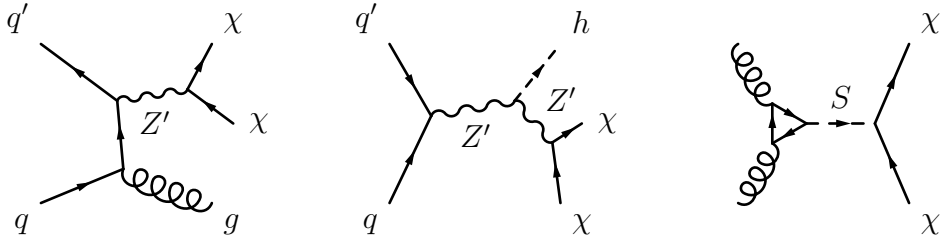

We implement the resonant production ($pp \rightarrow X$) and production with one jet ($pp \rightarrow X + j$) followed by decay of the mediator $X$ into DM fermions.  For simplicity, DM fermions are assumed to be Dirac. The vector boson model ($\mathrm{Z}^\prime$) allows vector and axial couplings to both SM fermions as well as DM.  Analogously, the scalar mediator ($\mathrm{S}$) model allows scalar and pseudoscalar couplings. The dark photon model (where SM couplings are determined by kinetic mixing with SM Z boson) can be accessed by setting the mixing parameter epsilon instead of setting each coupling individually.   In the case of the $\mathrm{Z}^\prime$ model, we also implement mono-higgs production where the coupling to the $\mathrm{Z}^\prime$ to the higgs can be set independently.  The diagrams corresponding to the production processes are shown in figure~\ref{fig:s-chan}.  

\begin{figure}[ht]
\begin{center}
\includegraphics[scale=0.5]{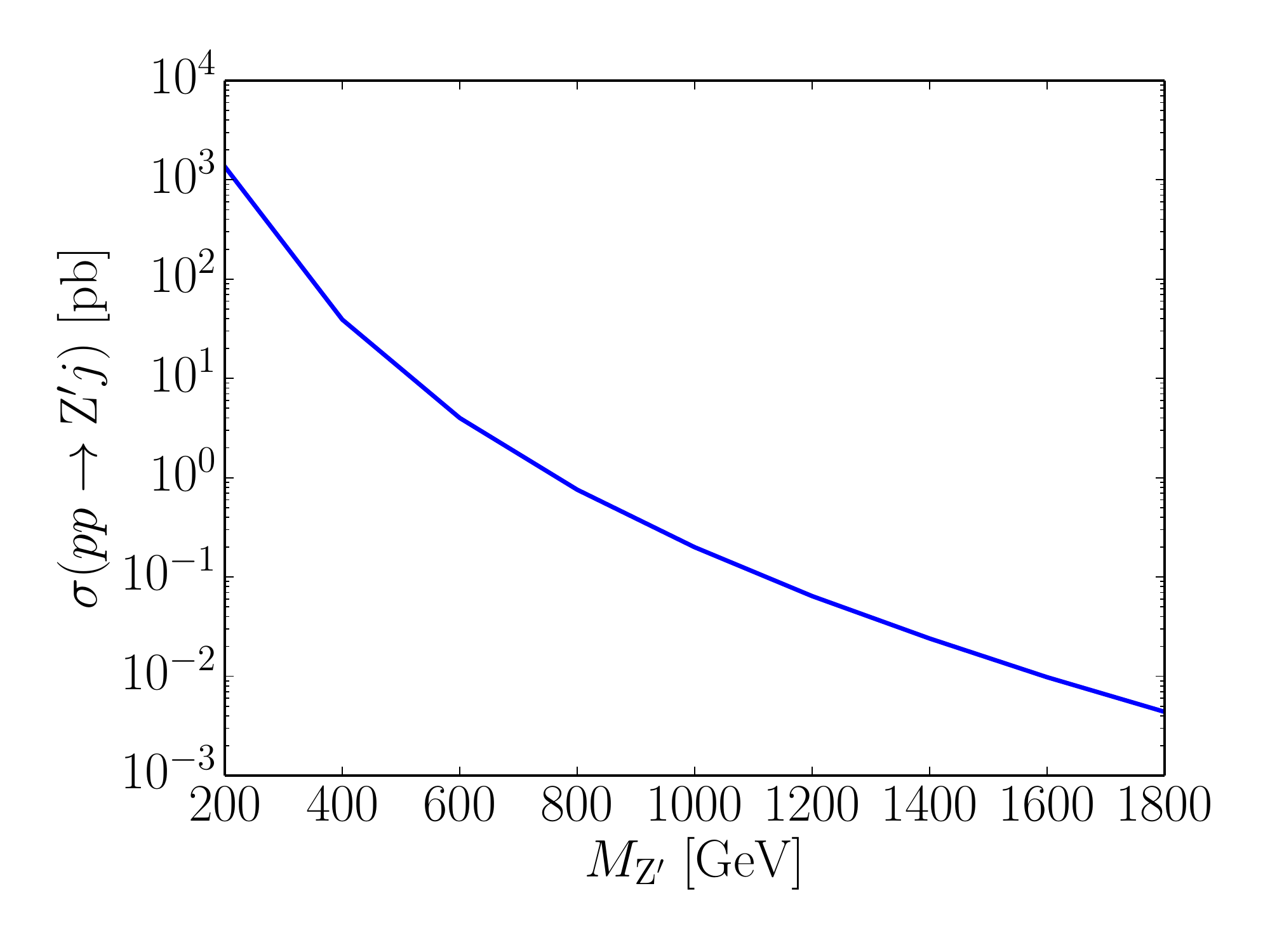}
\end{center}
\caption{\label{fig:axial} Mono-jet production cross section for the pure axial vector $\mathrm{Z}^\prime$ with $g_{\mathrm{Z}^\prime} = 0.1$ and $m_\chi = 10 $ GeV. }
\end{figure}

The relevant terms of the Lagrangian are given below. Table~\ref{tab:params} lists the descriptions of all relevant parameters implemented in the code.

\begin{equation}
\mathcal{L}_{Z^\prime} \supset -\frac{1}{4} Z^{\prime \mu \nu} Z^\prime_{\mu \nu} - M_{Z^\prime}^2 Z^{\prime \mu} Z^\prime_\nu + g_{Z^\prime} \bar f \gamma_\mu (v_f + a_f \gamma_5 ) f Z^\prime_\mu
\label{eqn:vector}
\end{equation}

\begin{equation}
\mathcal{L}_S \supset |\partial S|^2 - \frac{1}{2} m_S^2 |S|^2 - m_f \bar f (v_f + a_f \gamma_5) f  S
\label{eqn:scalar}
\end{equation}

%

\subsection{Scalar t-channel mediator}

This scalar mediator model is inspired by the simplest co-annihilation assumption where a dark matter fermion achieves the right relic density by co-annihilating with a charged scalar.  The mediator is assumed to have quantum numbers of a right-handed lepton (except spin) and the flavour of which can be set by choosing which SM lepton it couples to.  It is also possible to have lepton flavour violation by choosing multiple non-zero Yukawas.  Primary production at the LHC is Drell-Yan, followed by decay into lepton and DM.  The lagranigian given by equation~\ref{eqn:slep} also assumes a $Z_2$ symmetry under which $\ell$ and the dark matter $\chi$ are odd.

\begin{equation}
\mathcal{L} \supset |D_\mu \tilde \ell|^2 - \frac{1}{2} m_{\tilde \ell}^2 |\tilde \ell|^2 +  \bar \chi ( \gamma^\mu \partial_\mu - m_\chi) \chi + (y_i  \bar \chi \ell^R_i \tilde \ell + \mathrm{h.c.} )
\label{eqn:slep}
\end{equation}

\begin{figure}[t]
\begin{center}
\includegraphics[scale=0.5]{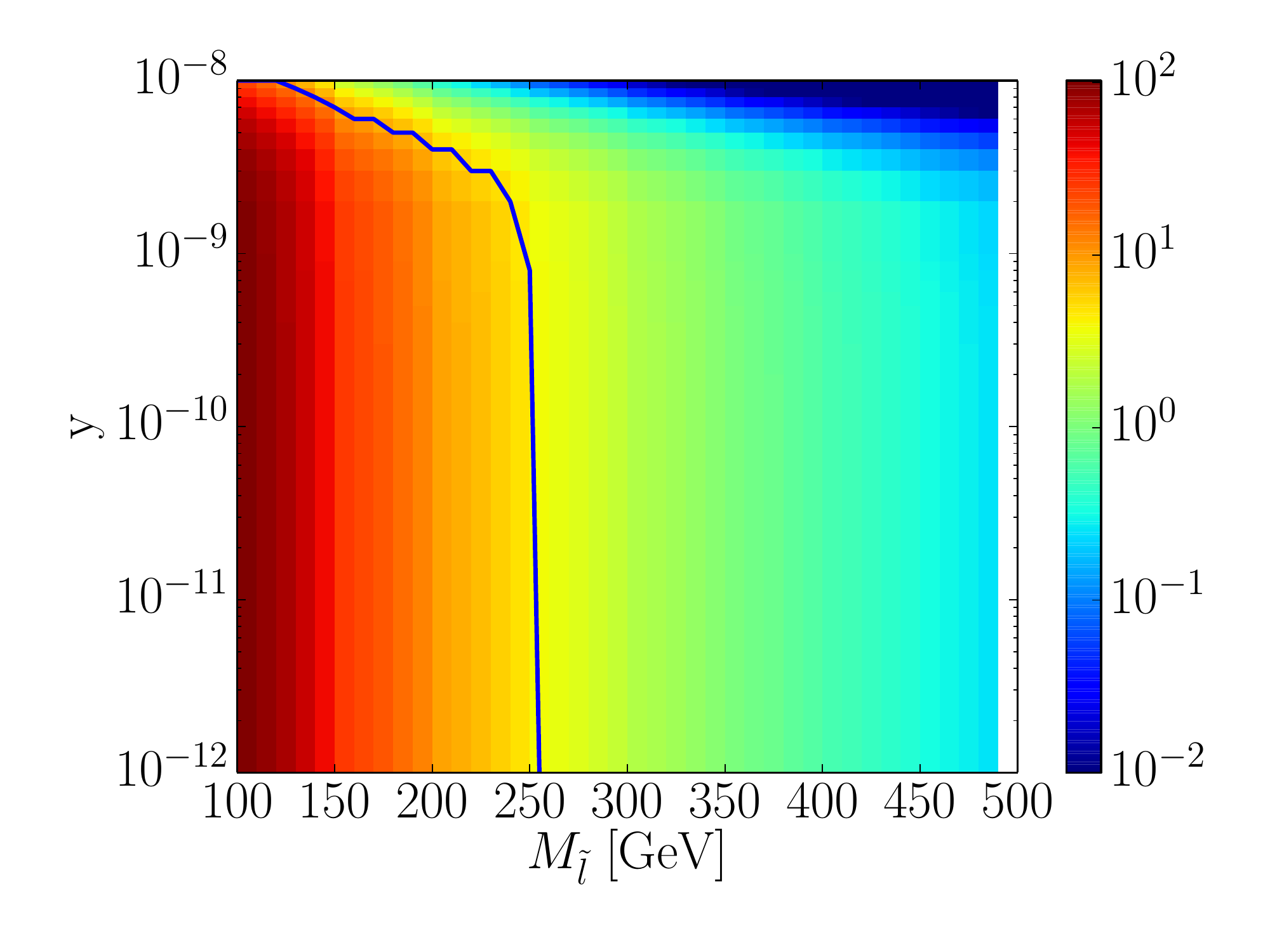}
\end{center}
\caption{\label{fig:slep} Limits from CMS charged particle search on the t-channel mediator model with only coupling to the right-handed electron. The colour bar shows the cross section of cross section of producing at least one particle per event that exits the CMS tracker.}
\end{figure}

This simple model can be used to model two cases where dark matter relic density is satisfied --- either by thermal freeze-out by co-annihilation or via freeze-in.  The first of these requires Yukawa couplings of order one and dark matter and scalar masses of order 100 GeV.  In case of freeze-in, imposing the right relic density forces the dark matter mass to be very light and the Yukawa coupling to be very small ($\lesssim 10^{-7}$), in effect giving a long-lived charged scalar on the scales of LHC detectors.  The reinterpretation of exclusions from CMS long-lived charged particle search can be seen in figure~\ref{fig:slep}.

\subsection{SU(2) n-plet mixed with singlet}

This model is a generalisation of the ``minimal dark matter" by adding an extra singlet fermion.  The minimal dark matter idea postulates that DM is the neutral component of a SU(2) $N$-plet, forbidden to decay by the symmetries of the SM.  The smallest $n$ was found to be 5.  However, mixed N-plets occur also in other theories like SUSY where dark matter can be a mixture of singlet, doublet and triplet states.  The model implemented here generalises this idea by defining a ``next-to-minimal" scenario of an $N$-plet mixed with a singlet ($N=2, 3, 5$).  The doublet scenario implemented here is a simplified version of the singlet-doublet model~\cite{Cohen:2011ec}, with both doublets assumed to have the same mixing.  The 5-plet scenario has one Dirac doubly charged fermion, one Dirac singly charged fermion and two neutral Majorana fermions. The 3-plet scenario has one Dirac singly charged fermion and two neutral Majorana fermions (similar to Bino-Wino scenarios of supersymmetry\cite{Nagata:2015pra}).  This model can be used as a benchmark to study all track-based long-lived signatures --- disappearing tracks, long-lived charges tracks (both singly and doubly charged), and displaced leptons.  Displaced vertex signatures can be obtained by looking at decays of the neutral DM partner.  A detailed description of the model and phenomenological study of the cases N=3 and 5 can be found in~\cite{Bharucha:2017ltz, Bharucha:2018pfu}.  The Lagrangian is given by
\begin{eqnarray}
\mathcal{L} & \supset &  i \psi^\dag_i  (\bar \sigma^\mu D_\mu) \psi_i + i\chi^\dag_j  \sigma^\mu \partial_\mu \chi_j - \frac{1}{2} (M_1 \chi \chi + M_2 \psi \psi)  \\
\mathcal{L}_\mathrm{mix;N=2} & = &  \lambda \psi^\dag \phi \chi + \mathrm{h.c.} \\
\mathcal{L}_\mathrm{mix;N=3} & = &  \frac{\lambda}{\Lambda} \psi^{\dag a} (\phi^\dag t^a \phi) \chi + \mathrm{h.c.} \\
\mathcal{L}_\mathrm{mix;N=5} & = &  \frac{\lambda}{\Lambda^3} \psi^{\dag a} (\phi \phi \phi \phi)^a \chi + \mathrm{h.c.}
\label{eqn:5plet}
\end{eqnarray}
where the term $(\phi \phi \phi \phi)^a$ is a schematic notation taken to mean a 5-plet combination from four higgs doublets. For the singlet-doublet~\cite{Cohen:2011ec} case, there is no scale suppression and the scale paramater \ttt{DM:Lambda} is translated into the mixing as $\lambda = (1~\mathrm{GeV}) / \Lambda$.  In the two other cases, as there is no way to separate effects of $\lambda$ and $\Lambda$, we simply keep $\lambda=1$ and encapsulate any effect of changing it into the value of $\Lambda$.  This allows us to describe all cases with just three parameters. In each case, the neutral mixing matrix and masses of the neutral eigenstates is calculated based on the parameters $M_1, M_2$ and $\Lambda$.  The parameter \ttt{DM:Nplet} allows choosing the value of $N$ for the N-plet.  

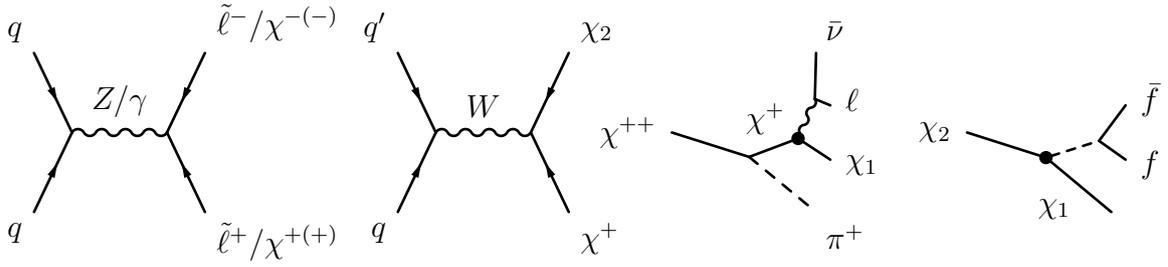
\begin{figure}[t]
\vskip 0.2in
\begin{center}
\begin{fmffile}{fiveplet}
\begin{fmfgraph*}(80,60) 
\fmfset{arrow_len}{2mm}
\fmfleft{i1,i2}
\fmfright{o1,o2}
\fmflabel{$\tilde \ell^+ / \chi^{+(+)}$}{o1}
\fmflabel{$\tilde \ell^- / \chi^{-(-)}$}{o2}
\fmflabel{$q$}{i1}
\fmflabel{$q$}{i2}
\fmf{fermion}{i1,v1}
\fmf{fermion}{i2,v1}
\fmf{photon, tension=0.8, label=$Z/\gamma$}{v1,v2}
\fmf{fermion}{o1,v2}
\fmf{fermion}{o2,v2}
\end{fmfgraph*}
\hspace{1.7cm}
\begin{fmfgraph*}(80,60) 
\fmfset{arrow_len}{2mm}
\fmfleft{i1,i2}
\fmfright{o1,o2}
\fmflabel{$\chi^{+}$}{o1}
\fmflabel{$\chi_2$}{o2}
\fmflabel{$q$}{i1}
\fmflabel{$q^\prime$}{i2}
\fmf{fermion}{i1,v1}
\fmf{fermion}{i2,v1}
\fmf{photon, tension=0.8, label=$W$}{v1,v2}
\fmf{fermion}{o1,v2}
\fmf{fermion}{o2,v2}
\end{fmfgraph*}
\hspace{0.8cm}
\begin{fmfgraph*}(60,60) 
\fmfset{arrow_len}{0mm}
        \fmfleft{i}
        \fmflabel{$\chi^{++}$}{i}
        \fmfright{o1,o2,o3,o4}
        \fmf{fermion,tension=1.5}{i,v1}
        \fmf{scalar}{v1,o1}
        \fmf{fermion, label=$\chi^{+}$,l.side=left}{v1,v2}
        \fmf{photon}{v2,v3}
        \fmf{fermion}{v2,o2}
        \fmf{fermion}{o3,v3,o4}
        \fmflabel{$\pi^+$}{o1}
        \fmflabel{$\chi_1$}{o2}
        \fmflabel{$\ell$}{o3}
        \fmflabel{$\bar \nu$}{o4}   
        \fmfdot{v2}     
\end{fmfgraph*}
\hspace{1.5cm}
\begin{fmfgraph*}(60,60) 
\fmfset{arrow_len}{0mm}
        \fmfleft{i}
        \fmflabel{$\chi_2$}{i}
        \fmfright{o1,o2,o3,o4}
        \fmf{fermion,tension=1.5}{i,v1}
        \fmf{fermion, label=$\chi_1$}{v1,o1}
        \fmf{scalar}{v1,v2}
        \fmf{fermion}{o2,v2,o3}
        \fmflabel{$f$}{o2}
        \fmflabel{$\bar f$}{o3}       
        \fmfdot{v1}   
\end{fmfgraph*}
\end{fmffile}
\end{center}
\vskip 0.2in
\caption{\label{fig:disp} From left to right, ({\it a}\/) Drell-Yan production of $\ell^\pm$ or $\chi^{\pm(\pm)}$ pairs, ({\it b}\/) production of $\chi^{\pm} \chi_2$, ({\it c}\/) decay chain for $\chi^{++}$ resulting in displaced leptons, and ({\it d}\/) decay of $\chi_2$ resulting in displaced vertex signatures. }
\end{figure}

The charged partners in this model are produced via Drell-Yan processes, followed by decay depending on the mixing parameters.  In particular, the doubly charged partner in the 5-plet model decays only via $\chi^{++} \rightarrow \chi^{+} \pi^+$ with a lifetime $c \tau$ of about 0.6mm determined by the splitting between $\chi^{++}$ and $\chi^{+}$.  The $\pi^+$ is too soft to be visible and the further decay of the $\chi^{+}$ into leptons can give displaced lepton or displaced jet signatures.  In the pure triplet case, the decay $\chi^{+} \rightarrow \chi_2 \pi^+$ gives disappearing tracks.  In cases of small mixing ($\Lambda > 10$~TeV), it is also possible to get displaced vertices from the decays of the $\chi_2$.  The feynman diagrams for the production and decay resulting in these signatures are shown in figure~\ref{fig:disp}.

\section{Conclusion and outlook}

We present here four models that can be used to study the production of dark matter in the simplest extensions of the Standard Model.  Aside from the standard simplified models where the dark matter is accompanied by a new s-channel mediator, we include two models where the dark matter particle is accompanied by charged partners that may be produced via Drell-Yan production.  These can also be used to study production of long-lived particles and their signatures at the LHC including displaced leptons and vertices, heavy charged particles and kink and disappearing tracks.  Although the models are similar to supersymmetric slepton and gaugino sector, the couplings are allowed to be completely independent and are not constrained by supersymmetry.  In the future, we plan to include production and decay of heavy neutral leptons and associated production of charged and neutral partners.

\vskip 0.1in 

\noindent \textbf{Acknowledgements:} This work has been carried out thanks to the support of the OCEVU Labex (ANR-11-LABX-0060) and the A*MIDEX project (ANR-11-IDEX-0001-02) funded by the "Investissements d'Avenir" French government program managed by the ANR. I would like to thank A.~Bharucha and F.~Br\"ummer for comments on the manuscript.

\bibliographystyle{JHEP}
\bibliography{paper-py8dm}

\pagebreak

\section*{Appendices: Pythia8 implementation}

\appendix

\section{PDG codes for new particles}
\begin{table}[h]
\begin{center}
\begin{tabular}{|l|p{12cm}|}
\hline
\textbf{PDG code} & \textbf{Description} \\
\hline
51 & Scalar Dark Matter (unused in current implementation) \\
52 & Fermionic Dark matter ($\chi_1$)\\
53 & Vector Dark Matter (unused in current implementation) \\
54 & (Pseudo-) Scalar mediator (S) \\
55 & (Axial-) Vector mediator ($Z^\prime$) \\
56 & Charged Scalar partner ($\tilde \ell$)\\
57 & Singly charged partner ($\chi^{+}$) \\
58 & Neutral partner ($\chi_2$)\\
59 & Doubly charged partner($\chi^{++}$) \\
\hline
\end{tabular}
\caption{\label{tab:pdg} PDG codes assigned to new particles.}
\end{center}
\end{table}

\section{List of implemented processes}
\begin{table}[h]
\begin{center}
\begin{tabular}{|l|p{12cm}|}
\hline
\textbf{Flag} & \textbf{Description} \\
\hline
\ttt{DM:gg2S2XX} &  Resonant (pseudo-) scalar production \\
\ttt{DM:gg2S2XXj} & Scalar + 1 jet (mono-jet) \\
\ttt{DM:ffbar2Zp2XX} & Resonant (axial-) vector production \\
\ttt{DM:ffbar2Zp2XXj} & Vector + 1 jet (mono-jet) \\
\ttt{DM:ffbar2ZpH} & Mono-higgs production \\
\hline
\ttt{DM:qqbar2DY} & Drell-Yan production of charged partners; exact process can be selected via the \ttt{DM:DYtype}  switch to select between production of co-annihilation partners \\
& \ttt{DM:DYtype = } \\ 
& 1: scalar lepton \\
& 2: charged fermion \\
& 3: doubly charged fermion \\
& 4: pair-production of neutral and singly-charged partner \\
\hline
\end{tabular}
\caption{\label{tab:procs} Flags for production processes.}
\end{center}
\end{table}

\pagebreak

\section{Parameters}

\begin{table}[h!]
\begin{center}
\begin{tabular}{|l|p{12cm}|}
\hline
\textbf{Parameter} & \textbf{Description} \\
\hline
$\mathbf{Z}^\prime$ & \\
\ttt{Zp:gZp} & Gauge coupling for $Z^\prime$\\
\ttt{Zp:coupH} & Coupling of $Z^\prime$ to SM higgs\\
\ttt{Zp:vu}, \ttt{Zp:vd} & Vector couplings of up- and down-type quarks\\
\ttt{Zp:vl}, \ttt{Zp:vv} & Vector couplings of charged and neutral leptons\\
\ttt{Zp:au}, \ttt{Zp:ad} & Axial-vector couplings of up- and down-type quarks\\
\ttt{Zp:al}, \ttt{Zp:av} & Axial-vector couplings of charged and neutral lepton\\
\ttt{Zp:vx}, \ttt{Zp:aX} & Vector and Axial Vector couplings of DM fermion \\
\ttt{Zp:epsilon} & Kinetic mixing for hidden photon (overwrites couplings to SM fermions) \\
\hline
$\mathbf{S}$ &   (All fermion couplings are multiplied by fermion mass)\\
\ttt{Sdm:vf} & Scalar coupling to SM fermions \\
\ttt{Sdm:af} & Pseudo-scalar coupling to SM fermions \\
\ttt{Sdm:vX} &  Scalar coupling to DM fermion \\
\ttt{Sdm:aX} & Pseudo-scalar coupling to DM fermion \\
\hline
\end{tabular}
\caption{\label{tab:params} Parameters for s-channel mediator models.}
\end{center}
\end{table}

\begin{table}[h!]
\begin{center}
\begin{tabular}{|l|p{12cm}|}
\hline
\textbf{Parameter} & \textbf{Description} \\
\hline

\hline
\ttt{DM:yuk1} & Yukawa coupling to the RH electron \\
\ttt{DM:yuk2} & Yukawa coupling to the RH muon \\
\ttt{DM:yuk3} &Yukawa coupling to the RH tau lepton \\
\hline
\ttt{DM:M1} & Mass parameter for singlet\\
\ttt{DM:M2} & Mass parameter for n-plet \\
\ttt{DM:Lambda} & Suppression scale of mixing \\
\ttt{DM:Nplet} & Representation of the mixed SU(2) N-plet.  Takes values 2, 3 or 5. \\
\hline
\end{tabular}
\caption{\label{tab:params2} Parameters for simplified models giving long-lived signatures.}
\end{center}
\end{table}

\end{document}